\documentclass[aps,english,twocolumn]{revtex4}
\usepackage{mathrsfs}
\usepackage{graphicx}
\usepackage{amsmath, amssymb}
\usepackage{babel}

\begin{document}

\title{Anderson Impurity in Helical Metal}
\preprint{1}

\author{Xiao-Yong Feng$^{1,2}$, Wei-Qiang Chen$^{2}$, Jin-Hua Gao$^{2}$, Qiang-Hua Wang$^{3}$, and Fu-Chun Zhang$^{2}$}
\affiliation{ $^1$Department of Physics, The  Normal University of
Hangzhou, Hangzhou 310036, China\\
$^2$Department of Physics and Centre of Theoretical and Computational Physics,\\
The University of Hong Kong, Hong Kong, China\\
$^3$National Laboratory of Solid State Microstructures and
Department of Physics,\\ The Nanjing University, Nanjing 210093,
China}

\begin{abstract}
We use a trial wavefunction to study the spin-1/2 Kondo effect of a
helical metal on the surface of a 3-dimensional topological
insulator. While the impurity spin is quenched by conduction
electrons, the spin-spin correlation of the conduction electron and
impurity is strongly anisotropic in both spin and spatial spaces. As
a result of strong spin-orbit coupling, the out-of-plane component
of the impurity spin is found to be fully screened by the orbital
angular momentum of the conduction electrons.
\end{abstract}
\maketitle

On the surface of a 3-dimensional (3D) topological insulator,
massless helical Dirac fermions emerge.\cite{Fu1} Specific materials
of 3D topological insulators have recently been studied and
observed.\cite{Hsieh,Xia,zhang-nphys,Chen} Interesting properties of
these helical Dirac fermions have become a focus of recent
studies.\cite{Fu2,Qi,Law,Tanaka} In a helical metal, spins are
coupled to momenta, so that magnetic properties are expected to
be highly non-trivial.  Theoretical study on this aspect, however,
has so far been limited to the effect of a classical magnetic
impurity.\cite{magneticimp}

The low temperature property of a quantum spin-1/2 magnetic
impurity or an Anderson impurity in a conventional
metal\cite{Anderson} has been an interesting and important problem
for decades in condensed matter physics. The problem has been
studied by using the renormalization group\cite{Wilson}, the Bethe
ansatz\cite{Wiegmann,Andrei}, the
$1/N$-expansion\cite{Coleman,Gunnarsson}, and the conformal field
theory\cite{Affleck}. The phenomenon is a well known Kondo effect,
in which the impurity spin is completely screened by the
conduction electrons. Because of the coupling between spins and
momenta, the Kondo effect in a helical metal will be interesting to be
examined.  In this paper, we use a variational method to
address this problem. We show that the correlation of the impurity
spin and the conduction electron spin density is strongly
anisotropic in both spatial and spin spaces, in contrast to the
isotropic screening in a conventional metal. While the impurity
spin is quenched at low temperatures, similar to the usual Kondo
problem, the correlation between the impurity spin and the spin of the
conduction electrons in the helical metal is significantly reduced.
The reduction of the spin-spin correlation in the direction perpendicular to the
surface can be shown explicitly to be compensated by the screening of the
orbital angular momentum of the conduction electrons.
The possible experimental consequences will be briefly discussed.

\begin{figure}[h]
 \includegraphics [width=8cm]{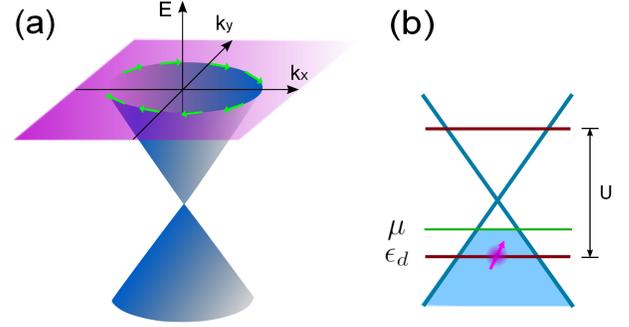}
 \caption{(color online). (a) Schematic energy-momentum dispersion in a helical metal.
Electron spin (green arrow) is perpendicular to momentum. (b) Energy diagram of a helical metal with
a singly occupied Anderson impurity state at energy $\epsilon_d$
in the absence of their coupling $H_{mix}$ in Eq. (\ref{Hamiltonian}). $\mu$ is the chemical potential.}
 \label{helicalmetal}
 \end{figure}

We consider a spin-1/2 magnetic impurity in a helical metal in a 2D
$x$-$y$ plane, described by the Hamiltonian,
\begin{eqnarray}\label{Hamiltonian}
&H &= H_{c} + H_{mix}+ H_{d}\\
&H_{c}&=\sum_{\mathbf{k}}c^{\dag}_{\mathbf{k}} [\hbar v_F  (\boldsymbol{\sigma}\times\mathbf{k})\cdot\hat{z}-\mu]c_{\mathbf{k}} \nonumber \\
&H_{mix}&=\sum_{\mathbf{k}}V_{\mathbf{k}}c^{\dag}_{\mathbf{k}}d+ h.c. \nonumber \\
&H_{d}&=(\epsilon_{d}-\mu)(n_{d\uparrow}+n_{d\downarrow})+Un_{d\uparrow}n_{d\downarrow}.
\nonumber
\end{eqnarray}
In the above equations, $c_{\mathbf{k}}$ and $d$ are annihilation
operators of the conduction electron with momentum $\mathbf{k}$ and
of the $d$-electron at impurity site in the spinor
representation, respectively. $\boldsymbol{\sigma}$ are the Pauli
matrices, $\mu$ is the chemical potential,
 and
$n_{d\sigma}=d^{\dagger}_{\sigma}d_{\sigma}$.
 $H_{c}$ describes a helical metal, and $H_d$ is a local impurity Hamiltonian with $\epsilon_d$
the impurity energy level and $U$ the Coulomb repulsion energy of
the two $d$-electrons on the same impurity site. $H_{mix}$ is a
hybridization term between the helical metal and the impurity state.

To start with, let us discuss the simple case $H_{mix}=0$ first. In
this limit, the helical metal and the impurity state are decoupled.
The single electron eigen-energy and its corresponding eigenstate of
the helical metal $H_c$ are given by
\begin{eqnarray}
\epsilon_{\mathbf{k} \pm} &=& \pm \hbar v_F |\mathbf{k}|-\mu \nonumber \\
\gamma_{\mathbf{k}
\pm}&=&\frac{1}{\sqrt{2}}(e^{\frac{i}{2}\theta_{\mathbf{k}}}c_{\mathbf{k}\uparrow}\pm
ie^{-\frac{i}{2}\theta_{\mathbf{k}}}c_{\mathbf{k}\downarrow})
\end{eqnarray}
with $\theta_{\mathbf{k}}$ the angle of the momentum $\mathbf{k}$
with respect to the $x$-axis that $\tan{\theta_{\mathbf{k}}}=k_y/k_x$
and "$\pm$" refer to upper and lower bands, respectively. For each
single particle state, its spin lies in the plane and is
perpendicular to the direction of its momentum.  Note that such a spin-momentum
relation has recently
been reported in a
spin- and angle-resolved photoemission spectroscopy (spin-ARPES)
experiment \cite{sARPES}. The ground state of $H_c$ is then given by
\begin{eqnarray}
|\Psi_0\rangle=\prod_{\{\mathbf{k}\pm
\}\in\Omega}\gamma^{\dag}_{\mathbf{k}\pm}|0\rangle, \nonumber
\end{eqnarray}
 where the
product runs over all the states within the Fermi sea $\Omega$. As
for the impurity part, we shall consider an interesting case, where
$\epsilon_d < \mu < \epsilon_d +U$, so that the impurity site is
singly occupied and has a local moment. The total energy of the
system $H_0 = H_c +H_d$ is then
\begin{eqnarray}
E_0 = \epsilon_d-\mu +
\sum_{\{\mathbf{k}s\}}\epsilon_{\mathbf{k}s}, \nonumber
\end{eqnarray}
 where $s=\pm$ is the
band index and hereafter the sum of $\{\mathbf{k}s\}$ is always over
the Fermi sea. In Fig. \ref{helicalmetal}, we illustrate the
electron dispersion and the spin of the Dirac fermion and the ground
state of a helical metal, in the absence of the hybridization with the impurity state.

To study the ground state of $H$ in the presence of $H_{mix}$, we
will use a trial wavefunction approach. Such a trial wavefunction
method was used to study the ground state of the Anderson impurity
problem in the conventional metal\cite{Varma, Gunnarsson} or in an
antiferromagnet\cite{Varma2}. For simplicity, we consider here the
large U limit, and exclude the double d-electron occupation at the
impurity site. We expect the qualitative physics should apply to the
case of finite but large U. The trial wavefunction for the ground
state is,
\begin{eqnarray}
|\Psi\rangle=(a_{0}+\sum_{\{\mathbf{k}s\}}
a_{\mathbf{k}s}d^{\dag}_{\mathbf{k}s}\gamma_{\mathbf{k}s})|\Psi_0\rangle
\end{eqnarray}
where
$d_{\mathbf{k}\pm}=\frac{1}{\sqrt{2}}(e^{\frac{i}{2}\theta_{\mathbf{k}}}
d_{\uparrow}\pm
ie^{-\frac{i}{2}\theta_{\mathbf{k}}}d_{\downarrow})$, and $a_{0}$
and $a_{\mathbf{k}s}$ are variational parameters
. They are to be
determined by optimizing the ground state energy. Note that by
construction, the impurity state is either singly occupied or
non-occupied.

The energy of $H$ in the variational state $ |\Psi \rangle$ is given
by
\begin{eqnarray}
E=\frac{\sum_{\{\mathbf{k}s\}}(E_0-
\epsilon_{\mathbf{k}s})a_{\mathbf{k}s}^2+
2V_{\mathbf{k}}a_{0}a_{\mathbf{k}s}+
\epsilon_{\mathbf{k}s}a_0^2}{a_0^2 +
\sum_{\{\mathbf{k}s\}}a_{\mathbf{k}s}^2}
\end{eqnarray}
The variational method leads to the equations below,
\begin{eqnarray}
(E-\sum_{\{\mathbf{k}s\}}\epsilon_{\mathbf{k}s})a_0&=&\sum_{\{\mathbf{k}s\}}V_{\mathbf{k}}a_{\mathbf{k}s}\\
(E-E_0+\epsilon_{\mathbf{k}s})a_{\mathbf{k}s}&=&V_{\mathbf{k}}a_0\label{a_k}
\end{eqnarray}
which may determine the binding energy $\Delta_{b}\equiv E_0 -E $
due to the hybridization,
\begin{eqnarray}\label{E2}
\epsilon_{d}-\mu-\Delta_b =
\sum_{\{\mathbf{k}s\}}\frac{|V_{\mathbf{k}}|^2}{\epsilon_{\mathbf{k}s}-\Delta_{b}}.
\end{eqnarray}
If $\Delta_{b}>0$, the hybridized state is stable against the
decoupled state. To proceed further, we introduce an energy cutoff
$\Lambda$ for the helical metallic state in our analysis, and
consider $V_{\mathbf{k}}=V\Theta(\Lambda-\hbar v_{F}|\mathbf{k}|)$,
where $\Theta(x)$ is a step function, which is 1 for $x>0$ and 0 for
$x<0$. The low energy physics is expected to be insensitive to the
cutoff $\Lambda$. A natural energy cutoff for the helical metallic
state in the topological insulator is the half of the bulk gap.\cite{Chen}
Eq.(\ref{E2}) enables us to determine the ground state energy $E$ or
the binding energy $\Delta_b$ and the ground state wavefunction. The
results are that the hybridization always leads to a binding
$\Delta_b >0$, and a magnetic screening of the conduction
electrons to the impurity spin at any $\mu \neq 0$. At the Dirac
point $\mu=0$, the binding and screening only occur at the
dimensionless hybridization strength $\alpha =\frac{2\pi
V^2}{\hbar^2 v_F^2}$ above a critical value. We will come back to
this point later.

\begin{figure}[h]
\includegraphics [width=6cm]{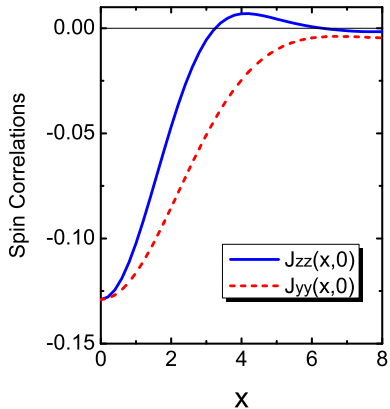}
\caption{(color online). Spatial Spin-spin correlation functions
$J_{yy}(x,0)$ and $J_{zz}(x,0)$ ($J_{xx}(x,0)=J_{zz}(x,0)$) in Eq. (8) as
functions of $x$, for a set of parameters $\Delta_{b}=0.005\Lambda$
and $\mu=-0.01\Lambda$. The length unit is $\hbar v_F/\Lambda$.
Different parameters show qualitatively similar feature as long as
$\Delta_b>0$.} \label{Jxyz}
 \end{figure}

We shall first examine the magnetic properties of the system, which
are most interesting. The central quantity we will examine is the
correlation function of the impurity spin
$\mathbf{S_d}=\frac{1}{2}d^{\dagger}\boldsymbol{\sigma}d$ at the impurity
site (set to be at the origin $\mathbf{r}=0$) and the conduction
electron spin density
$\mathbf{S_c}=\frac{1}{2}c^{\dagger}(\mathbf{r})\boldsymbol{\sigma}c(\mathbf{r})$
 in the ground state, namely
\begin{eqnarray}
J_{uv}(\mathbf{r})\equiv \langle
S_{c}^{u}(\mathbf{r})S_{d}^{v}(0)\rangle,
\end{eqnarray}
where $\langle Q \rangle$ is the ground state average of $Q$, and
$u,v=x,y,z$ are the spin indices. If $u$ and $v$ are both on the
$x-y$ plane, we then have, by the rotational symmetry,
$J_{uv}(\mathbf{r})=J_{u'v'}(\mathbf{r'})$, if $u' =R_z(\beta)u$,
$v' =R_z(\beta)v$, and $\mathbf{r'}= R_z(\beta) \mathbf{r}$, with
$R_z(\beta)$ a rotational operator of angle $\beta$ along the
z-axis. $J_{\mu,\nu}(\mathbf{r})$ can be calculated by using the
trial wave-function in Eq. (3). Note that the variational parameter
$a_{\mathbf{k}\pm}$ is independent of $\theta_{\mathbf{k}}$ since it
can be expressed as $Va_{0}/(\pm\hbar
v_{F}|\mathbf{k}|-\mu-\Delta_{b})$ according to the Eq. (\ref{a_k}).
In terms of $a_{\mathbf{k}\pm}$, the diagonal components are found
to be
\begin{eqnarray}
J_{zz}(\mathbf{r}) &=& -\frac{1}{8}|\mathcal{B}(\mathbf{r})|^2+\frac{1}{8}|\mathcal{A}(\mathbf{r})|^2\nonumber \\
J_{xx}(\mathbf{r})&=& -\frac{1}{8}|\mathcal{B}(\mathbf{r})|^2-
\frac{1}{8}Re(\mathcal{A}^2(\mathbf{r})) \nonumber\\
J_{yy}(\mathbf{r})&=&-\frac{1}{8}|\mathcal{B}(\mathbf{r})|^2+
\frac{1}{8}Re(\mathcal{A}^2(\mathbf{r})) \nonumber
\end{eqnarray}
where $\mathcal{A}(\mathbf{r})=\sum_{\{\mathbf{k}s\}}s
e^{i(\mathbf{k}\cdot\mathbf{r}+\theta_{\mathbf{k}})}a_{\mathbf{k}s}
$ and
$\mathcal{B}(\mathbf{r})=\sum_{\{\mathbf{k}s\}}e^{i\mathbf{k}\cdot\mathbf{r}}a_{\mathbf{k}s}$.
$J_{zz}(\mathbf{r})=J_{zz}(r)$ is rotational symmetric around the
impurity site, while $J_{xx}(\mathbf{r})$ and $J_{yy}(\mathbf{r})$
are highly anisotropic in space. In Fig. \ref{Jxyz}, we show
$J_{zz}(r)$ and $J_{yy}(x,0)$.  Note that $J_{xx}(x,0) =J_{zz}(x,0)$
and their signs oscillate in space. A negative (positive) value
means anti-parallel (parallel) correlation between the impurity spin
and the conduction electron spin density. Due to the absence of the
spin $SU(2)$ symmetry, the off-diagonal components are not zero in
general and they are
\begin{eqnarray}
J_{xy}(\mathbf{r})&=&J_{yx}(\mathbf{r})=-\frac{1}{8}Im(\mathcal{A}(\mathbf{r})^2), \nonumber \\
J_{xz}(\mathbf{r})&=&-J_{zx}(\mathbf{r})=\frac{1}{4}Im(\mathcal{B}(\mathbf{r})\mathcal{A}(\mathbf{r})), \nonumber \\
J_{yz}(\mathbf{r})&=&-J_{zy}(\mathbf{r})=-\frac{1}{4}Re(\mathcal{B}(\mathbf{r})\mathcal{A}(\mathbf{r})).
\nonumber
\end{eqnarray}

\begin{figure}[h]
 \includegraphics [width=8cm]{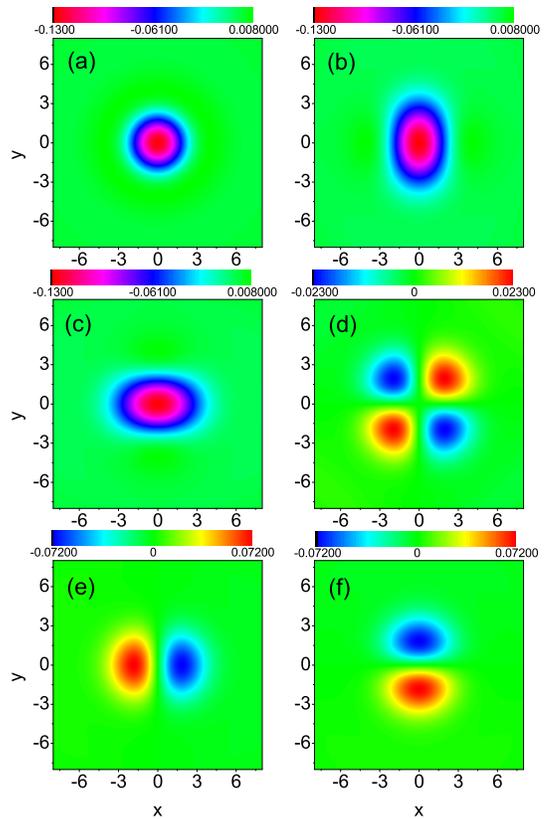}
 \caption{(color online). Spatial spin-spin correlation functions plotted in x-y plane.
(a)$J_{zz}(x,y)$; (b)$J_{xx}(x,y)$; (c)$J_{yy}(x,y)$;
(d)$J_{xy}(x,y)$ or $J_{yx}(x,y)$; (e)$J_{xz}(x,y)$ or
$-J_{zx}(x,y)$); (f)$J_{yz}(x,y)$ or $-J_{zy}(x,y)$. The parameters
are the same as those in Fig. \ref{Jxyz}.}
 \label{JJ}
 \end{figure}

The spatial distributions of the spin correlations in 2D are plotted
in Fig. \ref{JJ}. As we can see, the magnitude of
$J_{yy}(\mathbf{r})$ is larger at the space
point $\mathbf{r}$ near the $x$-axis than near the $y$-axis, and the
magnitude of
$J_{xx}(\mathbf{r})$ is larger at the space
point $\mathbf{r}$ near the $y$-axis than near the $x$-axis.

The anisotropic spin correlation function implies that the Kondo
screening to a magnetic impurity in a helical metal has more complex
texture. In the strong spin-orbit coupled system, the screening to the impuriy spin may be contributed
from both the spin and orbital angular momentum of the conduction electrons.
In what follows, we will show that in comparison with the Kondo effect in a conventional metal,
the spin-spin screening is largly reduced, and the z-component of the impurity spin  is fully screened by the
orbital angular momentum.  Let us first examine the correlation between the total spin of
the conduction electron, $\mathbf{S_c}=\int \mathbf{S_c(\mathbf{r})}
d^2\mathbf{r} $
 and the impurity spin. We define
\begin{eqnarray}
I_{u}= \langle S_{c}^{u}S_{d}^{u}\rangle/\sum_{\{\mathbf{k}s\}}a_{\mathbf{k}s}^2, \nonumber
\end{eqnarray}
with $u=x,y,z$, where the denominator is the occupation number of
the d-electron. $I=\sum_{u}I_{u}$ is a measure of spin-spin
screening strength. $I=0$ if the conduction electron and impurity
state are decoupled. $I=-3/4$ for a spin-1/2 magnetic impurity in a
conventional metal, corresponding to a spin singlet of two spin-1/2.
In the present helical metal, when the Fermi level is at or below
the Dirac point, we find $I_{x}=I_{y}=-\frac{1}{8}$ and $ I_{z}=0$,
therefore $ I=-\frac{1}{4}$,
which is one third of the value of $-3/4$ in the usual Kondo
problem.  Note that although $J_{zz}$ is non-zero locally, its overall
contribution to the spin-spin screening is zero. The spin-spin
screening comes from the in-plane spin components. Since  $I$ is
independent of any parameters including the energy cutoff (the only
requirement is $\mu < 0$), we expect that this is a general property of the
helical metal. When the Fermi level is above the Dirac point, there
is a deviation from the above result. For instant, with the
parameters $\mu=0.01\Lambda$ and $\Delta_{b}=0.005\Lambda$ the
deviation is about sixty percents of $-1/4$. At $\mu >0$, the upper
Dirac cone is partially filled. The electron state in the upper
Dirac cone has spin orientation opposite to the corresponding state
in the lower Dirac cone, which provides a stronger screening than
induced by the lower band only. We remark that the difference
between $\mu<0$ and $\mu>0$ is due to the particle-hole asymmetry in
our model parameter with U large. If $U+2(\epsilon_d-\mu)=0$, we
have particle-hole symmetry, and we expect correponding symmetries
for $\mu<0$ and $\mu>0$.

The conduction electron spin is not a good quantum number in a
helical metal.  To better understand the Kondo screening in the
helical metal, we shall consider contribution from the orbital
angular momentum.  Because of the two dimensionality, only z-component of
the orbital angular momentum can be considered here. The system is
invariant with respect to a simultaneous rotation of both spin and
space along the z-axis at the origin of the impurity site.
Therefore, the z-component of the total angular momentum
$J^z=L^z+S_c^z+S_d^z$ commutes with the Hamiltonian H and is a good
quantum number, where $L^z$ is the total orbital angular moment of
the conduction electrons. Since the ground state preserves the time
reversal symmetry, we have $J^z|\Psi\rangle=0$. Because of this and
$\langle S_c^zS_d^z\rangle=0$ discussed above, we have
\begin{eqnarray}
\frac{\langle
L^zS_d^z\rangle}{\sum_{\{\mathbf{k}s\}}a_{\mathbf{k}s}^2}=\frac{\langle
J^zS_d^z-S_c^zS_d^z-(S_d^z)^2\rangle}{\sum_{\{\mathbf{k}s\}}a_{\mathbf{k}s}^2}=-\frac{1}{4}.
\end{eqnarray}
Therefore, although spin-spin screening in the z-direction in the
present Kondo problem vanishes, the orbital angular momentum $L^z$
replaces the role of the conduction electron spin $S_c^z$
to screen the impurity spin.

The spatial anisotropic correlations may be detected in spin
resolved scanning tunneling spectroscope (STM) experiments. The
magnetic susceptibility is finite in this variational theory. In the
light of the unconventional spin correlations, exactly how the
susceptibility behaves has to be answered by more elaborate
treatments such as quantum Monte Carlo and numerical renormalization
group. This issue is left for further investigations.

\begin{figure}[h]
 \includegraphics [width=8.5cm]{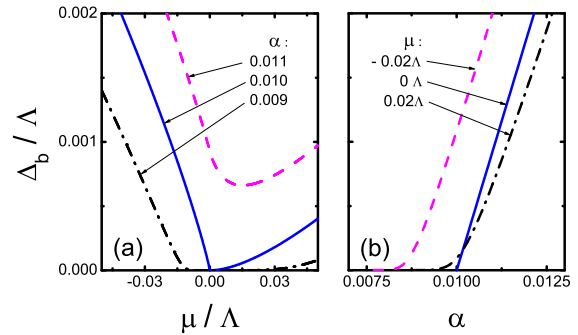}
 \caption{(color online). Binding energy $\Delta_{b}$ as functions of the chemical potential $\mu$ and
 the hybridization strength $\alpha$. $\Lambda$ is an energy cutoff.}
 \label{delta}
 \end{figure}

We now discuss the binding energy $\Delta_b$ of the system. From Eq.
(\ref{E2}), we obtain
\begin{eqnarray}
&&\epsilon_{d}-\mu-\Delta_{b}+\alpha(\Lambda-|\mu|)\nonumber\\
&=&\alpha(\mu+\Delta_{b})\ln{\frac{4(\Lambda+\mu+\Delta_{b})\Delta_{b}}{(\mu+|\mu|+2\Delta_{b})^2}}
\end{eqnarray}
In the limit of small $\alpha$, and $\alpha\Lambda<
\mu-\epsilon_{d}$, we have
\begin{eqnarray}
\Delta_{b}
&\approx\Lambda\exp{\left[-\frac{\mu-\epsilon_{d}-\alpha\Lambda}{\alpha|\mu|}\right]}&, \mu <0 \nonumber \\
\Delta_{b}
&\approx\frac{\mu^2}{\Lambda}\exp{\left[-\frac{\mu-\epsilon_{d}-\alpha\Lambda}{\alpha|\mu|}\right]}
&, \mu>0.
\end{eqnarray}
If the Fermi level is at the Dirac point, $\mu=0$, $\Delta_{b}$ has
a positive value solution only if $\alpha >
\alpha_{c}=|\epsilon_{d}|/\Lambda$. In other words, for a given
hybridization strength $\alpha$, there is a critical value of the
impurity $d$-level, below which there is no magnetic screening. This
is due to the vanishing density of states at the Dirac point. This
result is similar to the impurity problem in graphene.
\cite{Sengupta,Zhuang}.  The binding energy as functions of the
chemical potential $\mu$ and the hybridization $\alpha$ are plotted
in Fig. \ref{delta}.

$\Delta_{b}$ is asymmetric about the point $\mu=0$. The solid curves
in Fig. \ref{delta} are for $\alpha=\alpha_{c}$ and $\mu=0$
respectively. At $\alpha>\alpha_{c}$, the hybridized state is always
stable. At $\alpha<\alpha_{c}$, the binding energy is strongly
reduced around the Dirac point and vanishes at $\mu=0$.  As $\alpha
\rightarrow \alpha_c +0^+$,
$\Delta_{b}\approx|\alpha-\alpha_{c}|\Lambda$. This behavior is
consistent with the result by using large-degeneracy method.\cite{Withoff}

In summary, we have examined a spin-1/2 Anderson impurity in a 2D
helical metal. The momentum-dependent spin orientation in the
helical metal shows interesting magnetic properties. We have used a
trial wavefunction method to study the system at the large Coulomb U
limit. While the impurity spin is quenched by the conduction
electrons, we find strong spin- and spatial-anisotropies in the
correlations between the impurity spin and the conduction electron
spin density. Because of the strong spin-orbit coupling, the orbital
angular momentum also contributes significantly to the screening of
the impurity spin. In particular, the out-of-plane component of the
impurity spin is  found to be fully screened by the orbital angular
momentum of the conduction electrons. After our paper was submitted,
there was a report on the experimental realization of the magnetic
doping on the surface of the 3D topological insulator Bi2Se3
\cite{Cha}, and a e-print by Zitko, who studied the essentially the
same model as ours and mapped it onto a conventional Ander- son
pseudogap impurity model.\cite{Zitko} He showed that the impurity
spin is fully screened. Our trial wave-function method reported here
is consistent with his result.

Finally we comment on the similarity and difference of the helical
metal with the graphene.  In terms of the eigen-energy problem, a
helical metal is often considered to be a quarter of a graphene,
where the pseudospin plays the similar role with the spin in helical
metal. The graphene has a two-fold spin degeneracy and two-fold
degeneracy associated with the two sublattices. Our results for the
binding energy are similar to the graphene. However, because of the
difference between the spin in the helical metal and the pseudospin
in graphene, the magnetic properties of the two are different.
If we consider only one of the two Dirac cones is coupled to the
spin-1/2 impurity, the graphene behaves like a conventional metal.

We acknowledge useful discussions with Yi Zhou and Naoto Nagaosa,
and acknowledge partial support from HKSAR and RGC. QHW was
supported by NSFC (under the Grant Nos.10974086 and 10734120) and
the Ministry of Science and Technology of China (under the Grant No.
2006CB921802).

\end{document}